\documentclass[12pt]{article}

\usepackage[margin=1in]{geometry}
\usepackage{amsmath, amssymb}
\usepackage[margin=1in]{geometry}
\usepackage{amsmath, amssymb}
\usepackage{graphicx}
\usepackage{hyperref}
\usepackage{booktabs}
\usepackage{enumitem}
\usepackage{parskip}
\usepackage[numbers]{natbib}
\usepackage{float}
\usepackage{subcaption}
\graphicspath{{~/dev/dev-papers/ImageArchive/journals/2026/ChronoRay}{../../../../local-image-archive/journals/2026/ChronoRay}}

\hypersetup{
    colorlinks = true,
    linkcolor  = blue,
    citecolor  = blue,
    urlcolor   = blue
}

\title{
    \textbf{Chrono::Ray: A Distributed Framework for \\
    High-Throughput Simulation-Based Analysis of Multibody Systems}
}

\author{
    Khailanii Slaton\textsuperscript{1,2} \and
    Dan Negrut\textsuperscript{1} \\
    \small \textsuperscript{1}Department of Mechanical \& Aerospace Engineering,
    University of Wisconsin-Madison \\
    \small \textsuperscript{2} Department of Electrical \& Computer Engineering,
    University of Wisconsin-Madison
}

\date{}

\begin{document}

\maketitle

\begin{abstract}

Large-scale simulation studies can provide invaluable insights across computational engineering
efforts, but they are often computationally demanding, requiring the use of
distributed computing, which is itself not a simple task.
Chrono::Ray addresses this challenge by integrating the high-fidelity multibody
dynamics simulation engine Chrono with the open-source distributed computing platform Ray.
The result is a modular workflow framework providing user-friendly abstractions
for large-scale engineering simulation studies, supporting scalable orchestration 
of large ensembles of simulation trials without requiring users to directly manage distributed infrastructure.
The current capabilities of the framework are demonstrated through two
representative examples: parameter recovery for a multibody lunar lander model, 
and design of experiments for parameters of a continuum terramechanics model.
Chrono::Ray is a part of the larger Project Chrono ecosystem and is released as
an open-source software package, with source code available at \url{https://github.com/uwsbel/chrono-ray.git}.

\end{abstract}

\section{Motivation}
\label{sec:motivation}

Computer simulation plays a crucial role in modern engineering design,
existing as a powerful tool for system development, validation, and
performance evaluation prior to physical prototyping. Specifically, high-fidelity
simulation environments allow engineers to analyze the performance and
behavior of complex systems under a wide range of operating conditions.

However, simulation-based analysis is inherently affected by two
complementary sources of uncertainty. \emph{Epistemic uncertainty}
arises from incomplete knowledge of model parameters or system behavior, 
possibly reduced by gathering additional information.
In simulation, this manifests as situations where the analyst knows the model is imperfect but
lacks sufficient data to fully constrain it. \emph{Aleatoric uncertainty}
is associated with irreducible stochastic variability in physical
processes, and cannot be eliminated through additional measurement or
modeling effort. In simulation, this can manifest as variability in initial
conditions, manufacturing tolerances, or environmental disturbances. 

To address these uncertainties, robust system analysis must be employed.
This typically involves many simulation trials in which model 
parameters are varied and the system is evaluated with respect to some common 
objective. For computationally intensive multibody models, these large-scale 
simulation efforts can quickly become prohibitively expensive when 
executed sequentially. As a result, scalable workflow orchestration and
distributed execution have become increasingly important components of
modern computational engineering and scientific computing pipelines.

To address this computational bottleneck, we introduce Chrono::Ray,
a distributed simulation framework consisting of the integration of the
high-fidelity multibody dynamics engine Chrono and the Ray distributed
computing platform. Rather than requiring users to directly manage distributed execution,
resource scheduling, or worker coordination, Chrono::Ray is a package of several modular, workflow-oriented 
interfaces that abstracts the complexity of distributed execution and provides
ready-made support for tasks such as parameter estimation, 
Bayesian inference, and robust design optimization. The resulting tool
brings large-scale statistical simulation within reach for computationally 
intensive simulation studies where sequential execution is no longer practical.

\section{Ray: A Distributed Computing Platform}
\label{sec:ray}

Ray~\cite{moritz2018} is an open-source distributed computing framework
designed to unify parallel and distributed execution under a single,
Python-native abstraction, allowing users to accelerate their workflows 
without expert knowledge of distributed computing. Originally developed at the University of
California, Berkeley, it is currently maintained as an active open-source
project\footnote{\url{https://github.com/ray-project/ray}} that has been 
commonly utilized in both academic research and industrial machine
learning efforts.

\subsection{Core Architecture}

At its foundation, Ray provides a \emph{task-based execution model} in
which arbitrary Python functions (decorated with \texttt{@ray.remote})
are dispatched to a distributed scheduler that manages resource
allocation, fault tolerance, and data movement transparently. Users need
not write explicit message-passing code or manage worker processes
directly; the framework handles all of this under the hood. Ray supports
execution on a single multicore workstation as well as on large
heterogeneous clusters, making it straightforward to scale a workflow
from a laptop to a cloud environment without modifying application code.

Ray exposes two main levels of API:

\begin{itemize}[leftmargin=*]
    \item \textbf{Ray Core} is the lower-level API that allows users to
    implement custom distributed workflows with more control over
    task parallelism, actor lifecycles, and shared object stores. 

    \item \textbf{Ray AIR (AI Runtime)} is a higher-level API that
    provides ready-made templates for common machine learning and
    optimization workflows. It is composed of several specialized
    libraries, including Ray Tune for hyperparameter search and Ray
    Train for distributed model training. Additionally, libraries in this 
    layer integrate seamlessly with the broader Python scientific computing 
    ecosystem (NumPy, SciPy, PyTorch, scikit-learn, and others).

\end{itemize}

\subsection{Ray Tune and Search Algorithms}

Of the libraries that compose Ray AIR, \emph{Ray Tune}~\cite{liaw2018tune}
is the most directly relevant to the engineering simulation workflows
targeted by this work. Originally designed for machine learning
hyperparameter optimization, Ray Tune's underlying execution model is
general enough to treat any parameterized Python callable as a
``trial,'' making it equally well-suited to driving large ensembles
of physics simulations. Its key components are as follows:

\begin{itemize}[leftmargin=*]

    \item \textbf{Search algorithms} determine how for each trial the
    values for parameters are selected. Available options include random
    search, grid search, Bayesian optimization (via BoTorch or
    Optuna), BOHB, and CMA-ES, among others.

    \item \textbf{Trial schedulers} manage the prioritization and, where
    applicable, early termination of trials. For simulation workloads that
    report intermediate metrics at runtime checkpoints, schedulers such as
    ASHA and Hyperband can terminate unpromising trials early to conserve
    compute resources.

    \item \textbf{Parameter space distributions} allow users to specify
    prior distributions over each parameter, including
    \texttt{Uniform}, \texttt{LogUniform}, \texttt{RandInt},
    \texttt{Randn}, and arbitrary grid-valued choices.

\end{itemize}

Because Ray Tune requires no modification to accept a Chrono simulation
function in place of a neural network training loop, it serves as the
natural backbone for the statistical workflows implemented in Chrono::Ray.

\section{Chrono::Ray: Framework Description}
\label{sec:framework}

Chrono::Ray is a modular software interface that bridges
PyChrono, the Python binding of the Project Chrono
multibody dynamics engine~\cite{tasora2016}, with the Ray distributed computing platform.
Its central design goal is to allow engineers and researchers to leverage
high-throughput computing (HTC) for engineering simulation studies with
minimal additional cognitive load. The main idea is that a user familiar with 
PyChrono only needs to allocate most of their effort to defining the simulation function 
and a parameter space. They then pass this information to their desired Chrono::Ray 
instance, in which the distributed execution, result aggregation, and interface to
the selected analysis workflow is handled.

\subsection{Design Philosophy}

The goal of Chrono::Ray is to provide user-friendly interfaces that handle most
of the complexity of the desired analysis workflow, requiring the user to know
little to nothing about distributed computing or Ray.

This is achieved by wrapping the relevant Ray internals within the framework
itself, so the user is never required to dig through Ray documentation directly.
Extensive inline documentation and per-class \texttt{info()} functions are
provided to guide the user through configuration and usage.

Each module follows a similar design pattern, in which the user mainly
interacts with a given workflow via the following two primitives:

\begin{itemize}[leftmargin=*]

    \item \textbf{\texttt{simulate\_fn(config)}} a user-defined PyChrono simulation
    script wrapped as a callable that accepts a \texttt{config} dictionary
    of sampled parameter values and returns a dictionary of floats to be 
    used transparently by the workflow's objective function. 

    \item \textbf{\texttt{param\_sample\_space : dict[str, ChR\_Distr]}}
    a dictionary mapping parameter names to sampling distributions defined
    in the \texttt{ChR\_Distr} (e.g., \texttt{uniform}, \texttt{loguniform},
    \texttt{choice}). The workflow's search algorithm samples parameter values 
    from these ranges transparently in order to configure and dispatch the 
    various simulation trials.  

\end{itemize}

These two primitives are passed to one of four user-facing workflow
classes, each of which adds its own configuration on top:

\begin{itemize}[leftmargin=*]

  \item \textbf{\texttt{ChRParamEst}} is a parameter estimation workflow
  that tunes simulation parameters to match a set of observed target values.
  It requires a \texttt{target\_sim\_outputs} dictionary of observed values
  to match and an \texttt{est\_rule} (e.g., least squares, MLE) that
  determines how each trial is scored. The objective function is constructed
  internally from these inputs.

  \item \textbf{\texttt{ChRBayesOpt}} is a Bayesian optimization workflow
  that searches for the parameter configuration that best optimizes a
  user-defined objective. It requires an \texttt{objective\_fn} that maps
  the raw simulation output to a scalar float and a \texttt{mode} of
  \texttt{"min"} or \texttt{"max"}. The search algorithm is fixed to
  Gaussian-process Bayesian optimization.

  \item \textbf{\texttt{ChROpt}} is a general-purpose optimization workflow
  that offers the same interface as \texttt{ChRBayesOpt} but with a
  pluggable search algorithm. The user may select any available Ray Tune
  search algorithm via \texttt{ChR\_SearchAlg}, which can be inspected
  at runtime via \texttt{ChR\_SearchAlg.info()}.

  \item \textbf{\texttt{ChRDoE}} is a design of experiments workflow that
  generates and runs a structured ensemble of parameter configurations in
  parallel. It requires a \texttt{sampling\_design} (full factorial, Latin
  hypercube, or Sobol sequence). There is no objective function; any outputs
  of interest must be logged internally within \texttt{simulate\_fn}.

\end{itemize}

All four workflow classes share the same two operating modes. In the
default mode (\texttt{FLAG\_auto\_run=True}), the workflow runs
immediately on construction with no further configuration possible.
In the manual mode (\texttt{FLAG\_auto\_run=False}), construction
stops after input validation, allowing the user to configure
optional settings (e.g., resources per trial, search algorithm
kwargs) via setters before calling \texttt{\_build()} and
\texttt{run()} explicitly.

Additionally, logging of outputs can be redirected from the console to a
timestamped \texttt{.txt} file in the working directory by setting
\texttt{FLAG\_log\_to\_file=True}, allowing the user to easily review
analysis results from Ray after the workflow has completed.

\begin{figure}[H]
    \centering
    \includegraphics[width=1.0\textwidth]{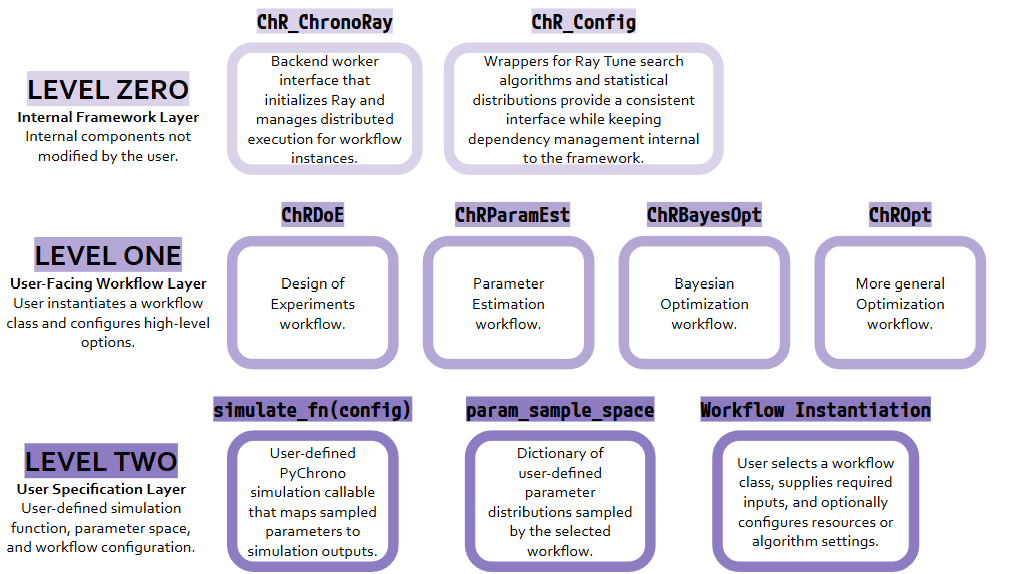}
    \caption{
      Chrono::Ray abstraction hierarchy showing user-defined inputs, workflow interfaces, and internal components for distributed execution.
      }
    \label{fig:overview}
\end{figure}

\section{Extensions}
\label{sec:workflows}

Beyond the workflows detailed above, there are several additional
simulation studies that the Chrono::Ray architecture is well-positioned
to support. 

\subsection{Uncertainty Quantification}

Uncertainty quantification (UQ) workflows propagate distributions over uncertain
input parameters through the simulation to characterize the resulting variability
in system outputs. The Chrono::Ray architecture is positioned to support Monte
Carlo and quasi-Monte Carlo sampling strategies, distributing the required
ensemble of runs across available workers and aggregating output statistics
such as mean, variance, and empirical confidence intervals.

\subsection{Sensitivity Analysis}

Sensitivity analysis workflows quantify the relative influence of each input
parameter on one or more outputs of interest. Variance-based Sobol sensitivity
indices~\cite{saltelli2010} are a natural fit, computed from a distributed
ensemble of simulation evaluations. Global sensitivity indices identify which
parameters drive the most output variance, helping to guide model simplification
and inform which parameters are worth calibrating or optimizing.

\subsection{Robust Design Optimization}

Robust design workflows search for parameter configurations that perform well
not just at a nominal operating point, but across a distribution of uncertain
or variable conditions. This is formulated as an optimization over a robustness
metric evaluated across an inner ensemble of Monte Carlo samples at each
candidate design. The Chrono::Ray architecture is positioned to parallelize
both the outer optimization loop and the inner ensemble evaluations.

\subsection{Control and Trajectory Optimization}

Control and trajectory optimization workflows treat controller gains, reference
trajectories, or motion profiles as the decision variables and use the simulation
to evaluate closed-loop performance. This includes PID gain tuning, open-loop
trajectory optimization, and model predictive control rollout evaluation.
The existing \texttt{simulate\_fn} and \texttt{objective\_fn} interface extends
naturally to this setting, with the parameter space simply defined over
controller or trajectory parameters rather than physical properties.

\subsection{Surrogate Modeling}

Surrogate modeling workflows use a distributed ensemble of simulation evaluations
to train a fast emulator, such as a Gaussian process, polynomial chaos expansion,
or neural network, that approximates the simulation input--output map. Once
trained, the surrogate can be queried in place of the full simulation for
downstream tasks such as optimization or sensitivity analysis.

\subsection{Multi-Fidelity and Adaptive Sampling}

Multi-fidelity workflows combine simulation evaluations at different levels of
model fidelity, such as varying integration step size or contact model
complexity, to reduce total computational cost. Adaptive sampling strategies
use early simulation results to focus additional evaluations in regions of the
parameter space where the objective or surrogate is most uncertain, improving
the information gained per unit of compute.

Taken together, these extensions illustrate the flexibility of the core
framework and its design principles, allowing a user to interact with a
common interface while leveraging different workflow instances to perform
a wide range of computational engineering studies.

\section{Results}
\label{sec:results}

The capabilities of Chrono::Ray are demonstrated through two
representative simulation studies. The first example considers parameter
recovery for a multibody lunar lander model, and is used to demonstrate the
parameter estimation workflow. The second example considers a design of
experiments study for a continuum terramechanics model, and is used to
demonstrate the ability of Chrono::Ray to coordinate structured ensembles of
high-fidelity simulations.

Together, these examples illustrate the primary design goal of the framework:
allowing users to define a simulation function and a parameter space, then
delegate the execution, sampling, and result handling to a workflow-oriented
interface.

\subsection{Parameter Recovery for a Multibody Lunar Lander}
\label{sec:results_lander}

The first example uses Chrono::Ray to perform parameter estimation for a
multibody lunar lander impact model. The lander model is implemented in
PyChrono and includes a custom honeycomb force model used to represent
energy absorption during landing. The goal of the parameter estimation study is
to recover model parameters that cause the simulated system response to match a
set of prescribed target outputs.

The estimated parameters are

\[
    \beta, \quad \alpha_2, \quad f_y,
\]

where $\alpha_2$  is the strut inclination angle, $\beta$ is the leg spread
angle, and $f_y$ is the yield force of the honeycomb crush model. Together,
these parameters control both the geometric configuration of the lander and
the force response of its energy-absorbing elements.

\[
    \texttt{peak\_accel},
    \qquad
    \texttt{energy\_absorbed}.
\]

These outputs are compared against target values using a least-squares
estimation rule. In this example, the user-facing Chrono::Ray workflow is
specified by defining the PyChrono simulation callable, the parameter sampling
space, the target simulation outputs, and the estimation rule. The resulting
parameter estimation problem is then executed through the \texttt{ChRParamEst}
interface.

\begin{figure}[H]
  \centering
  \includegraphics[width=0.85\textwidth]{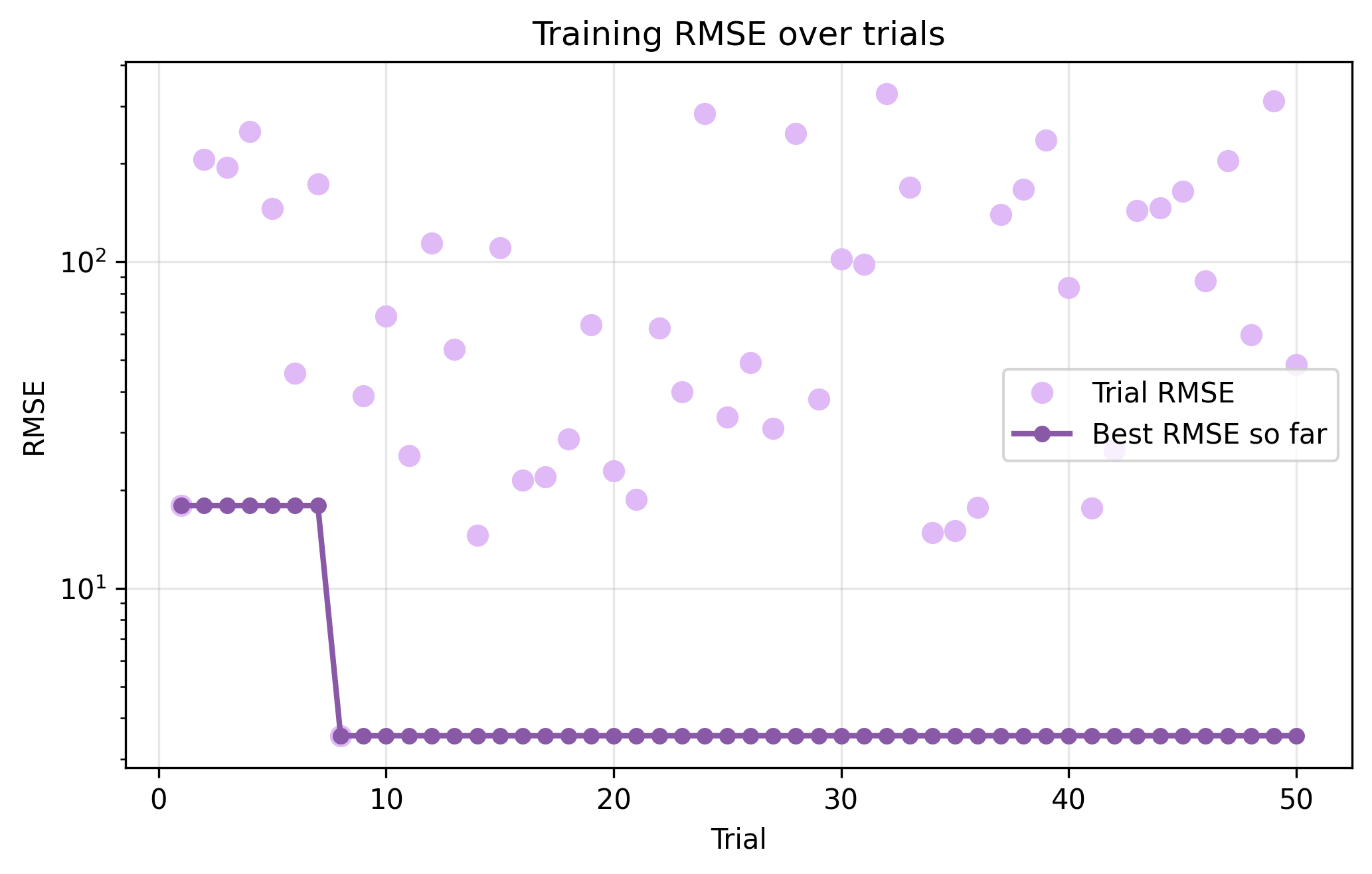}
  \caption{
  Parameter estimation results for the multibody lunar lander example.
  Individual markers show trial RMSE values, and the curve shows the
  best RMSE observed up to each trial.
  }
  \label{fig:lander_rmse_trials}
\end{figure}

For the representative run considered here, Chrono::Ray executed 50 candidate
parameter configurations on rigid terrain. The resulting RMSE values are shown in
Fig.~\ref{fig:lander_rmse_trials}, together with the best RMSE observed up to
each trial. The individual trial errors vary substantially across the sampled
parameter space, indicating that many candidate parameter configurations do not
reproduce the prescribed target response. However, the best-so-far curve drops
sharply at trial 8, where the best parameter configuration is identified. This
trial achieved a root mean squared error of approximately

\[
    \mathrm{RMSE} \approx 3.53,
\]

with a least-squares objective value of approximately

\[
    J \approx 623.56.
\]

After this trial, no later sampled configuration produced a lower error. This
behavior is expected for a search over a broad parameter space, where the goal
is not necessarily to produce monotonic improvement in every trial, but rather
to evaluate a collection of candidate configurations and retain the best one
observed.

The wall-clock time over the completed trials was approximately

\[
    28.66~\mathrm{s},
\]

for a multi-body system interacting with rigid terrain.

These results demonstrate the ability of Chrono::Ray to turn a PyChrono impact
model into a distributed parameter recovery workflow with minimal additional
user-side code. The user is not required to write explicit parallel execution
logic, manage workers, or manually aggregate trial outputs. Instead, the
parameter estimation study is expressed in terms of the simulation function,
the admissible parameter ranges, the target outputs, and the desired estimation
rule. Chrono::Ray then handles the repeated simulation calls, records the
trial-level results, and reports the best-performing parameter configuration.

\subsection{Design of Experiments for a Continuum Terramechanics Model}
\label{sec:results_doe}

The second example demonstrates the Chrono::Ray design of experiments workflow
using a continuum terramechanics simulation. The motivation behind this example 
is to gain insight into how the terrain parameters visually influence the 
simulated terrain behavior through an
angle-of-repose simulation setup in which an initially narrow column of granular
material collapses under gravity and spreads over a flat surface. The specific terramechanics 
model considered here is the Modified Cam-Clay rheology model for Chrono::CRM~\cite{unjhawala2025}. 

The design of experiments study varies six material and numerical parameters:

\[
    \mu_s, \quad \rho, \quad \lambda, \quad \kappa, \quad E, \quad \nu,
\]

where \(\mu_s\) is the static friction coefficient, \(\rho\) is the material
density, \(\lambda\) and \(\kappa\) are Modified Cam-Clay parameters, \(E\) is
Young's modulus, and \(\nu\) is Poisson's ratio. The sampled parameter ranges are

\[
    \mu_s \in [0.4, 1.2],
\]

\[
    \rho \in [1520, 1780],
\]

\[
    \lambda \in [0.02, 0.10],
\]

\[
    \kappa \in [0.005, 0.03],
\]

\[
    E \in [5\times 10^5, 5\times 10^6],
\]

\[
    \nu \in [0.2, 0.45].
\]

A Latin hypercube sampling design is used to generate 20 parameter
configurations over this six-dimensional parameter space. The study is executed
using the \texttt{ChRDoE} interface with a maximum of two concurrent trials.
For each sampled configuration, the simulation constructs the coupled
multibody-SPH system, assigns the sampled material properties, initializes the
particle column, advances the system forward in time, and writes particle data
to a run-specific output directory.

The Modified Cam-Clay parameters are also checked for physical admissibility.
Configurations for which

\[
    \kappa \geq \lambda
\]

are rejected before the simulation is run. This provides a simple example of how
problem-specific constraints can be embedded directly in the simulation
function while still allowing Chrono::Ray to manage the outer workflow.

Unlike the parameter estimation example, the design of experiments workflow does
not require an objective function. Instead, its purpose is to generate and
execute a structured ensemble of simulations whose outputs can be analyzed
afterward. In this case, each trial writes particle snapshots at the prescribed
output frequency, allowing the resulting pile geometry and flow behavior to be
post-processed outside the Chrono::Ray workflow.

This example demonstrates that Chrono::Ray is not limited to optimization or
parameter recovery tasks. It can also be used as a high-throughput simulation
driver for exploratory studies in which the goal is to sample a parameter space
and collect simulation outputs for later analysis. The user specifies the
sampling design, the number of trials, and the concurrency level, while the
framework handles the dispatch of individual simulation runs.

\begin{figure}[htbp]
  \centering

  \begin{subfigure}[t]{0.31\textwidth}
      \centering
      \includegraphics[width=\textwidth]{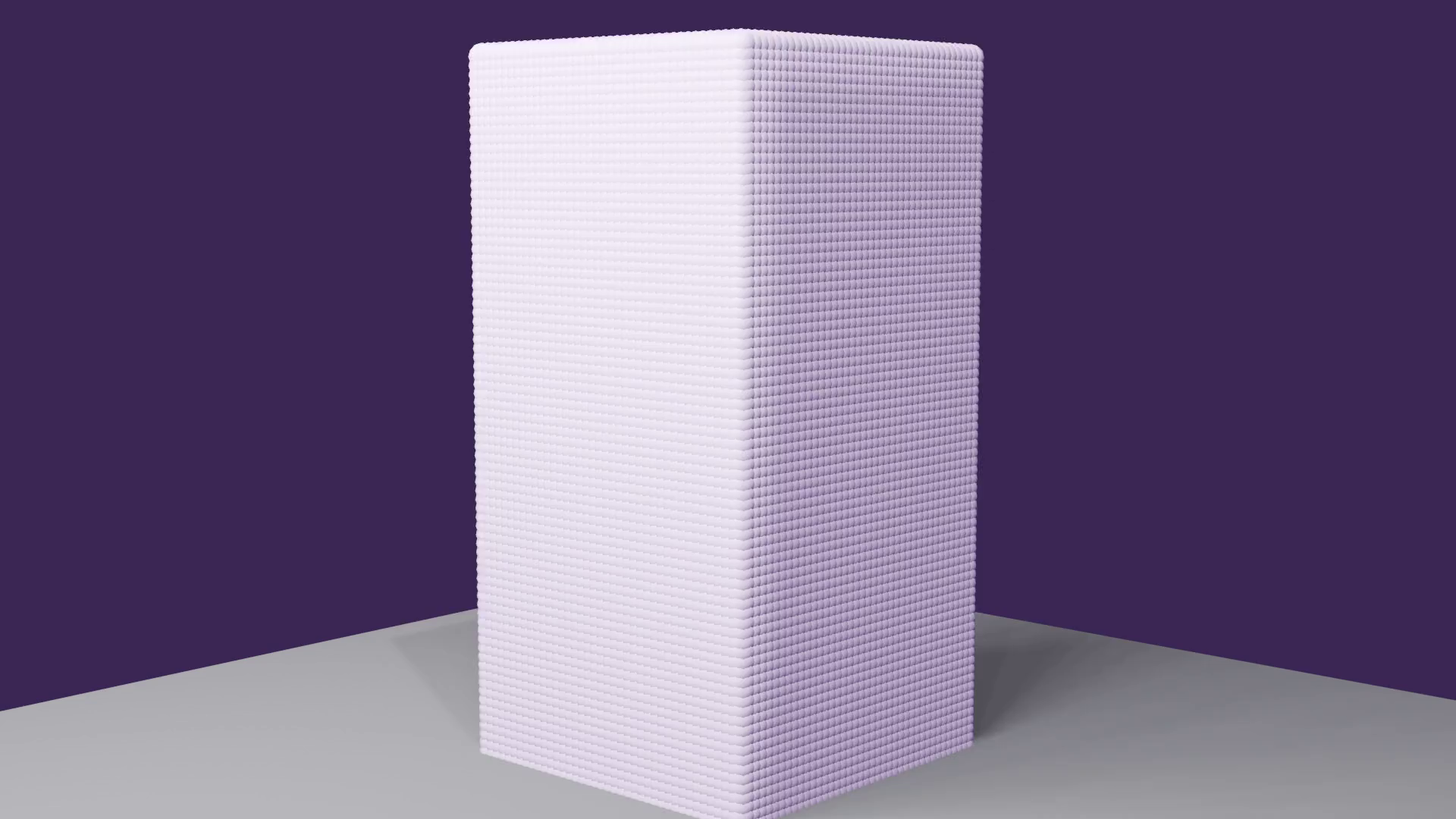}
      \caption{Beginning}
      \label{fig:mcc_beginning}
  \end{subfigure}
  \hfill
  \begin{subfigure}[t]{0.31\textwidth}
      \centering
      \includegraphics[width=\textwidth]{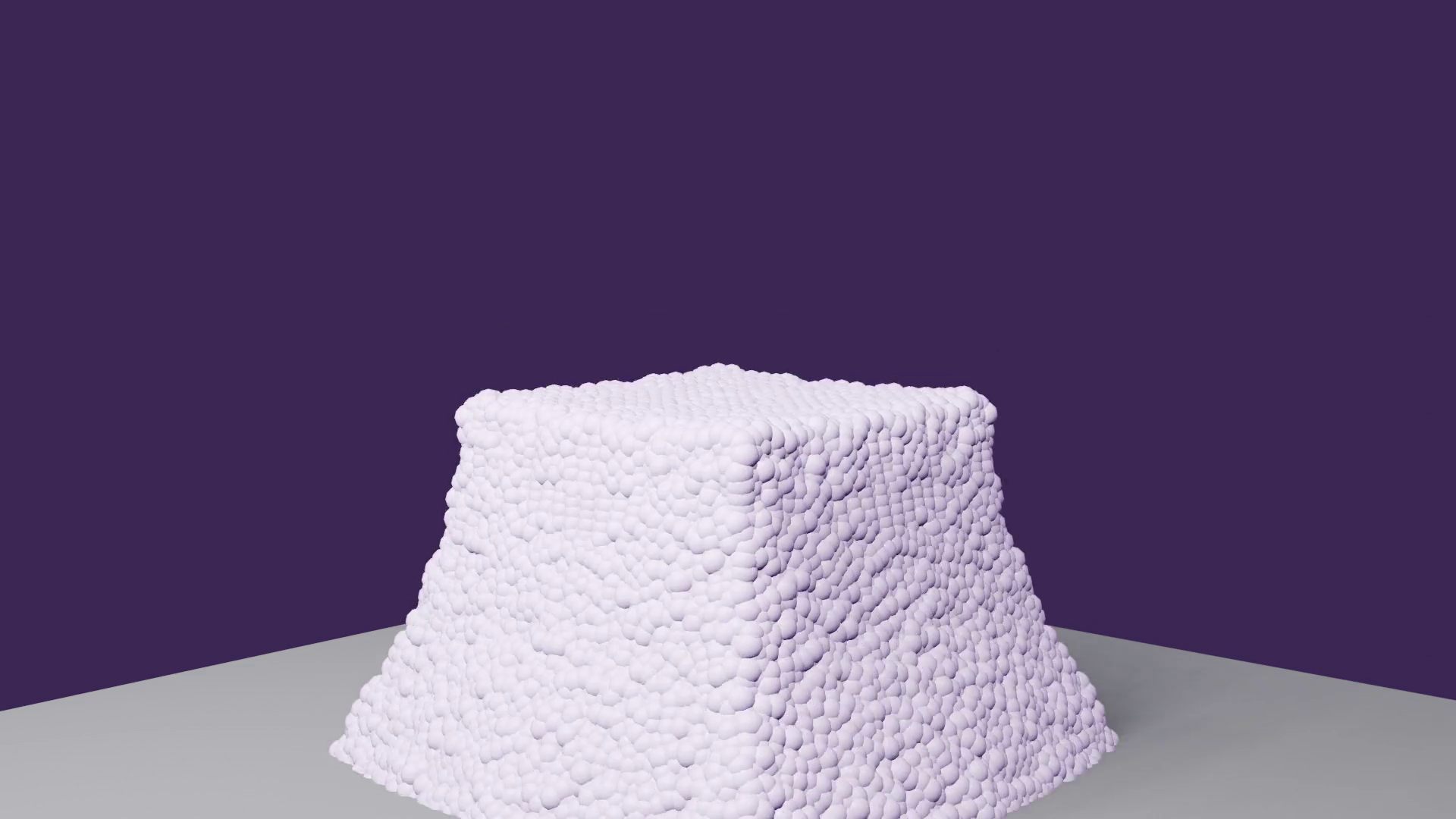}
      \caption{Middle}
      \label{fig:mcc_middle}
  \end{subfigure}
  \hfill
  \begin{subfigure}[t]{0.31\textwidth}
      \centering
      \includegraphics[width=\textwidth]{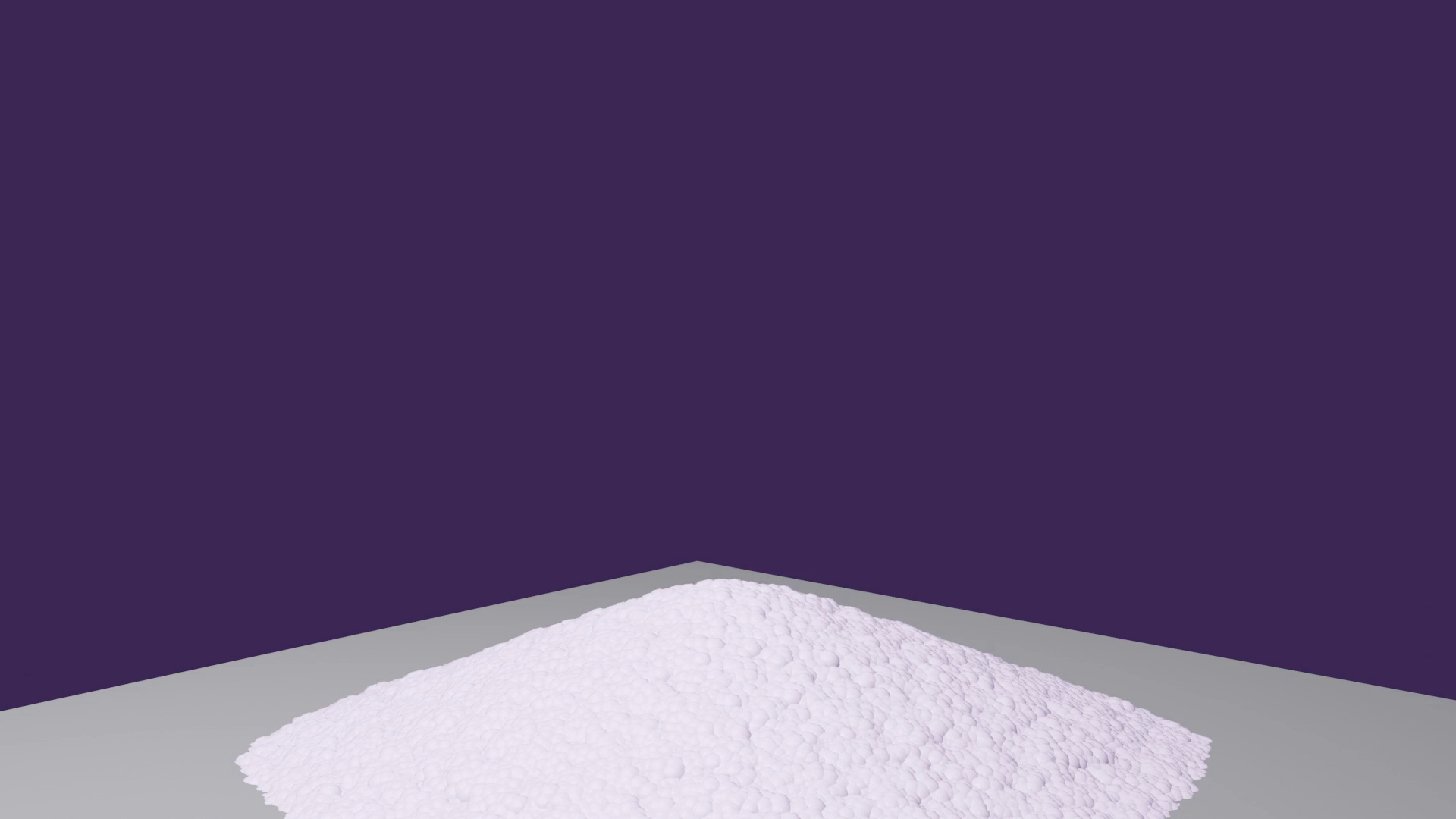}
      \caption{End}
      \label{fig:mcc_end}
  \end{subfigure}

  \caption{
  Representative snapshots from the continuum terramechanics design of
  experiments example, showing the material configuration at the beginning,
  middle, and end of one of the simulation trials rendered in Blender.
  }
  \label{fig:mcc_snapshots}
\end{figure}

\subsection{Discussion}
\label{sec:results_discussion}

The two examples exercise complementary parts of the Chrono::Ray interface.
The lunar lander example demonstrates a closed-loop parameter estimation
workflow, where each trial is scored against target outputs and the best
parameter configuration is reported. The terramechanics example demonstrates an
open-ended design of experiments workflow, where the primary goal is to generate
a distributed ensemble of simulation outputs rather than optimize a scalar
objective.

The common feature across both examples is that the user-facing structure of the
workflow remains essentially the same. In each case, the user defines a
simulation function, specifies a parameter space, and selects the appropriate
Chrono::Ray workflow class. This consistency is important because it allows
different statistical and computational studies to be expressed using the same
basic programming model. A simulation model developed for one workflow can
therefore be reused in another with minimal restructuring.

The examples also demonstrate the role of Chrono::Ray as an abstraction layer
between PyChrono and Ray. The framework does not replace the underlying
simulation model, nor does it require the user to rewrite the model in a
specialized distributed programming style. Instead, repeated simulation
execution is wrapped in workflow-specific interfaces that expose only the
configuration needed for the corresponding analysis task.

Together, these examples show how Chrono::Ray supports both objective-
driven and ensemble-driven computational engineering workflows. The parameter
estimation example applies when quantitative target values are available and
simulation outputs can be mapped to a scalar loss. The design of experiments
example applies when no scalar objective is prescribed and the desired outcome
is a collection of simulation data over a structured parameter space. These two
use cases represent common patterns in high-throughput simulation studies and
illustrate how Chrono::Ray reduces the overhead associated with executing large
numbers of high-fidelity simulations.

\section{Conclusions}
\label{sec:conclusions}

This work introduced Chrono::Ray, a distributed simulation framework that
integrates the Project Chrono multibody dynamics engine with the Ray distributed
computing platform. The framework provides workflow-oriented interfaces for
large-scale simulation studies, allowing users to express parameter estimation,
optimization, and design of experiments tasks in terms of a simulation function
and a parameter space rather than explicit distributed execution logic.

The representative examples demonstrate how Chrono::Ray can be used to perform
parameter recovery for a multibody lunar lander model and design of experiments
for a continuum terramechanics model. In both cases, the same basic programming
model is preserved: the user defines the PyChrono simulation, specifies the
parameters to be explored, and selects the corresponding Chrono::Ray workflow.
This consistency reduces the overhead associated with moving from a single
simulation model to a high-throughput computational study.

Chrono::Ray is intended to serve as a lightweight abstraction layer between
high-fidelity engineering simulation and modern distributed computing tools. By
hiding much of the complexity associated with task scheduling, worker
coordination, and result aggregation, the framework makes large-scale
simulation-based analysis more accessible to engineers and researchers working
with computationally intensive multibody models. The source code and 
examples are publicly available at \url{https://github.com/uwsbel/chrono-ray.git}.


\end{document}